\documentclass[aps,pra,reprint,superscriptaddress,longbibliography,nofootinbib]{revtex4-2}

\usepackage{amsmath,amssymb,amsthm,bm,mathtools}
\usepackage{graphicx}
\usepackage{hyperref}
\usepackage{physics}
\usepackage{microtype}
\usepackage{xcolor}
\usepackage{caption}
\usepackage{subcaption}
\usepackage{capt-of}
\usepackage{graphicx}

\hypersetup{colorlinks=true,citecolor=blue,linkcolor=blue,urlcolor=blue}

\newcommand{\ii}{\mathrm{i}}
\newcommand{\ee}{\mathrm{e}}
\renewcommand{\Tr}{\operatorname{Tr}}
\newcommand{\avg}[1]{\left\langle #1 \right\rangle}
\newcommand{\Eext}{E_{\mathrm{ext}}}
\newcommand{\crit}{\mathrm{crit}}

\newcommand{\Id}{\mathbb{I}}

\newcommand{\amend}[1]{\textcolor{black}{#1}}
\begin{document}

\title{Thermal screening and critical scaling of quantum energy teleportation in a harmonic chain}

\author{Taisanul Haque}
\email{taisanul.haque@stud.uni-goettingen.de}
\affiliation{Institute for Theoretical Physics, University of G\"ottingen, G\"ottingen, 37077, Germany}

\begin{abstract}
	We develop a finite-temperature Gaussian-state formulation of quantum energy teleportation in the one-dimensional harmonic chain. For Gibbs states, the optimized measurement-feedback protocol reduces to thermal two-point functions. In the single-site protocol, the extracted energy is governed by a single correlator, which makes the thermodynamic and near-critical limits analytically tractable. At fixed finite temperature, the extracted energy is exponentially screened with distance and is controlled by a thermal correlation length for both critical and non-critical $\alpha$. In the zero-temperature critical limit taken after the thermodynamic limit, we derive the exact asymptotic law $E_{\mathrm{ext}}\sim d^{-4}$. Numerical results confirm both regimes and resolve the crossover responsible for the apparent drift of the effective decay exponent at intermediate distances. We also analyze squeezed Gaussian measurements and show that they modify the extraction prefactor: $p-$squeezing enhances the extracted energy, whereas $q-$squeezing suppresses it, without changing the large-distance scaling.
\end{abstract}
\maketitle

\section{Introduction}

Quantum energy teleportation (QET) is a protocol in which a local measurement injects energy into a correlated many-body system, classical communication transfers the measurement outcome, and a distant conditional local operation extracts positive energy from another site \cite{Hotta2008distribution,Hotta2009spin,Hotta2010review}. The protocol does not transport usable energy during the classical-communication step. Instead, it uses local information together with pre-existing correlations of the state. In this way, QET links quantum information, quantum thermodynamics, and many-body physics.

A large part of the literature has focused on ground-state protocols \cite{Yusa2011QuantumHall,Hotta2013EntanglementBound,Hotta2014NoDistanceLimit,VerdonAkzam2016QuditQET,Ikeda2023CriticalityQFT,Ikeda2023SPTQET}. At finite temperature and in passive mixed states, however, detailed studies have mainly concentrated on few-body systems. In particular, Gibbs-state QET was analyzed for two spins by Frey, Gerlach, and Hotta \cite{Frey2013Gibbs} and for a three-spin Ising chain by Trevison and Hotta \cite{Trevison2015Gibbs}. \amend{Alongside these studies, our earlier work \cite{haque2024aspects} examined finite-temperature energy extraction in a minimal two-qubit XY model, focusing on its relation to entanglement, discord, and the passivity of the initial state.} This line of work clarified that nonzero temperature does not immediately destroy the protocol and that useful resources can survive in passive mixed states. Later developments connected QET with strong local passivity, including the original strong-local-passivity formulation \cite{Frey2014SLP}, the experimental activation of strong local passive states with quantum information \cite{RodriguezBriones2023SLPexp}, the construction of strong local passive states in the minimal QET model \cite{Wu2024SLPminimal}, and strong-QET protocols \cite{Fan2024StrongQET}. Recent reviews show that QET is now discussed in a broader quantum-thermodynamic and experimental context \cite{Ragula2025Review}.

QET has also been developed as an operational probe of correlated quantum matter. In particular, it has been linked to critical behavior in quantum field theory and to global and topological order, including symmetry-protected topological phases \cite{Ikeda2023CriticalityQFT,Ikeda2023SPTQET}. Related studies have explored feedback-induced entropy change in the Heisenberg chain, impurity effects in Kondo settings, and nonlocal correlations in the Kitaev spin liquid \cite{Itoh2023HeisenbergQET,Ikeda2023KondoQET,Matsueda2025MajoranaQET}.

\amend{The present work moves from these few-body spin settings to a spatially extended bosonic system, where both a controlled thermodynamic limit and analytically tractable long-distance behavior are available.} The one-dimensional harmonic chain is a natural model for this purpose. It has a gap-closing critical point, a well-defined thermodynamic limit, and Gaussian Gibbs states that permit an explicit treatment of both the state and the local operations. The same quadratic Hamiltonian also describes, after canonical rescaling, a chain of coupled linear microwave resonators, with node flux and charge playing the roles of canonical coordinate and momentum \cite{Houck2012}. In this language, the regime $\alpha\to 1$ corresponds to weak local pinning relative to intersite coupling, where the gap closes, the correlation length grows, and long-wavelength correlations become dominant.

Earlier work by Nambu and Hotta established ground-state QET in the harmonic chain and studied both single-site and block-measurement settings \cite{Nambu2010}. Their numerical analysis reported an apparent critical decay close to $\abs{\langle H_B\rangle}\sim 2\times10^{-3} d^{-3.6}$ at large distance, and no analytic derivation of that exponent was given \cite{Nambu2010}. The same work also explored enlarged measured regions on Alice's side, again at zero temperature and mainly numerically \cite{Nambu2010}.

The main results of the present work are as follows. For translationally invariant Gibbs states of the harmonic chain, the optimized Gaussian QET protocol reduces, in the single-site setting, to thermal two-point correlators. This reduction makes the ordered thermodynamic and critical limits analytically transparent. At fixed finite temperature, the extracted energy is exponentially suppressed at large distance and is controlled by a thermal correlation length. In the zero-temperature critical limit taken after the thermodynamic limit, the relevant coupling function admits a closed form and yields the asymptotic behavior $\Eext \sim d^{-4}$. We further show that squeezed Gaussian measurements modify the extraction prefactor but not the asymptotic scaling. In this way, the harmonic-chain QET protocol becomes an analytically controlled probe of thermal Gaussian correlations.

The paper is organized as follows. Section \ref{sec:model} introduces the harmonic-chain Hamiltonian, its physical realizations, and the two-point correlators of the Gibbs state.
Section \ref{sec:protocol} derives the optimized extracted energy for an arbitrary Gaussian block measurement and presents parameter surveys of the extracted energy [Fig. \ref{fig:Eext_alpha_beta_distance}]. Section \ref{sec:exact} specializes to the single-site protocol and establishes the key identity $J_q = h_d$. Section \ref{sec:asymptotics} derives the exact Matsubara series at finite temperature and the closed-form $d^{-4}$ law at zero temperature, with Figs. \ref{fig:hd_crititcal} and \ref{fig:Eext_benchmark} providing quantitative benchmarks.
Section \ref{sec:numerics} presents numerical diagnostics of the crossover between algebraic and exponential regimes [Fig. \ref{fig:Eext_distance_scaling}]. Section \ref{sec:squeezing} analyzes squeezed Gaussian measurements and shows that squeezing modifies the extraction efficiency without changing the large-distance scaling class [Figs. \ref{fig:fig5} and \ref{fig:sPOVM_loglog}].
Section \ref{sec:discussion} discusses the results and outlines open directions.

\section{Model and physical setting\label{sec:model}}

\subsection{Pinned phononic chain}

We begin from a standard spring-mass chain with onsite pinning,
\begin{equation}
H_{\mathrm{sm}}=\sum_j\left[\frac{P_j^2}{2M}+\frac{K_0}{2}u_j^2+\frac{K}{2}(u_{j+1}-u_j)^2\right]
\label{eq:Hsm}
\end{equation}
Expanding the elastic term gives
\begin{equation}
H_{\mathrm{sm}}=\sum_j\left[\frac{P_j^2}{2M}+\frac{K_0+2K}{2}u_j^2-Ku_ju_{j+1}\right]
\end{equation}
Define the characteristic frequency
\begin{equation}
\Omega_0:=\sqrt{\frac{K_0+2K}{M}}
\end{equation}
and dimensionless canonical variables
\begin{equation}
q_j:=\sqrt{M\Omega_0}\,u_j,
\quad
p_j:=\frac{P_j}{\sqrt{M\Omega_0}},
\quad [q_j,p_k]=\ii\delta_{jk}
\end{equation}
Then $$H_{\mathrm{sm}}=\frac{\Omega_0}{2}\sum_j\left[p_j^2+q_j^2-\alpha q_jq_{j+1}\right],
\quad
\alpha=\frac{2K}{K_0+2K}$$
Measuring energies in units of $\Omega_0$ yields the dimensionless Hamiltonian
\begin{equation}
H=\frac12\sum_{j=1}^N\left(p_j^2+q_j^2-\alpha q_jq_{j-1}\right)
\label{eq:Hdimless}
\end{equation}
with periodic boundary conditions. The parameter range $0<\alpha<1$ corresponds to a gapped pinned chain. The critical point $\alpha=1$ is the gap-closing limit.

\subsection{Microwave-resonator interpretation}

The same quadratic form arises in a chain of linear microwave resonators. For node fluxes $\Phi_j$ and conjugate charges $Q_j$, a capacitively or inductively coupled quadratic circuit has the generic form
\begin{equation}
H_{\mathrm{res}}=\sum_j\left[\frac{Q_j^2}{2C}+\frac{\Phi_j^2}{2L}+\frac{(\Phi_{j+1}-\Phi_j)^2}{2L_c}\right]
\label{eq:Hres}
\end{equation}
After the substitutions $u_j\leftrightarrow \Phi_j$, $P_j\leftrightarrow Q_j$, $M\leftrightarrow C$, $K_0\leftrightarrow 1/L$, and $K\leftrightarrow 1/L_c$, Eq. \eqref{eq:Hres} reduces by the same canonical rescaling to Eq. \eqref{eq:Hdimless}. The present QET setting can therefore be interpreted either as a universal bosonic low-energy theory or as an effective description of a circuit-QED bosonic lattice \cite{Houck2012}.

\subsection{Gibbs state and local energy}
The initial state is the Gibbs state
\begin{equation}
	\rho_\beta=\frac{\ee^{-\beta H}}{\Tr(\ee^{-\beta H})}
\end{equation}
\amend{For $0<\alpha<1$, $H$ is a positive quadratic Hamiltonian.
	A linear canonical transformation diagonalizes it into
	independent harmonic oscillators, whose thermal states are
	Gaussian. Since Gaussianity is preserved under linear
	canonical transformations, the Gibbs state $\rho_\beta$ is
	also Gaussian. Its first moments vanish, and the symmetrized
	$q$--$p$ covariance vanishes in equilibrium. Consequently,
	the state is completely specified by its $q$--$q$ and
	$p$--$p$ covariance functions. See, appendices \ref{app:A} and \ref{app:B} for complete derivations. In the thermodynamic limit,
	we denote these functions by}
\begin{align}
	g_n(\beta,\alpha)&=\langle q_j q_{j+n}\rangle
	\label{eq:gn_def}\\
	h_n(\beta,\alpha)&=\langle p_j p_{j+n}\rangle
	\label{eq:hn_def}
\end{align}
which take the integral forms
\begin{align}
	g_n(\beta,\alpha)
	&=\frac{1}{2\pi}\int_0^{2\pi}\frac{\dd\theta}{2\omega_\alpha(\theta)}
	\coth\qty(\frac{\beta\omega_\alpha(\theta)}{2})\cos(n\theta)
	\label{eq:gn_int}\\
	h_n(\beta,\alpha)
	&=\frac{1}{2\pi}\int_0^{2\pi}\frac{\omega_\alpha(\theta)}{2}
	\coth\qty(\frac{\beta\omega_\alpha(\theta)}{2})\cos(n\theta)\,\dd\theta
	\label{eq:hn_int}
\end{align}
with $\omega_\alpha(\theta)=\sqrt{1-\alpha\cos\theta}$.

\amend{Here $g_n$ and $h_n$ are, respectively, the position and
	momentum covariance functions at separation $n$. The
	nonlocal distance dependence of the single-site protocol is
	ultimately carried by $h_d$, while the onsite variances
	$g_0$ and $h_0$ determine the local noise and response
	prefactors. A normal-mode derivation of these covariance
	functions is given in Appendix \ref{app:B}.}

For later use we define the local Hamiltonian contribution probed by Bob's displacement,
\begin{equation}
H_B=\frac12 p_{n_B}^2+\frac12 q_{n_B}^2-\frac\alpha2 q_{n_B}(q_{n_B-1}+q_{n_B+1})-\frac\epsilon2
\label{eq:HB}
\end{equation}
where the constant shift $\epsilon$ cancels in the extracted-energy difference and will be left implicit. We also introduce
\begin{equation}
Q_B:=q_{n_B}-\frac\alpha2(q_{n_B-1}+q_{n_B+1})
\label{eq:QB}
\end{equation}

With the equilibrium correlators and the operator $H_B$ fixed, the QET problem reduces to how Alice's measurement noise and Bob's conditional displacement probe these thermal correlations. We now formulate the Gaussian measurement-and-feedback protocol for a general measured block.

\section{Gaussian QET protocol in the Gibbs state\label{sec:protocol}}

Alice measures a block $A$ of $m=2\ell+1$ sites with a coherent-state POVM. On each measured site she defines
\begin{equation}
\hat b=\sqrt{\frac{\Omega}{2}}\,\hat q+\frac{\ii}{\sqrt{2\Omega}}\,\hat p
\label{eq:bop}
\end{equation}
where $\Omega>0$ is the measurement width. The coherent states $\ket{X,P}$ satisfy $\hat b\ket{X,P}=c\ket{X,P}$ with $c=\sqrt{\Omega/2}\,X+\ii P/\sqrt{2\Omega}$. The block POVM is
\begin{equation}
\Pi_A(\tilde X,\tilde P)=\frac{1}{(2\pi)^m}\prod_{j=1}^m \ket{X_j,P_j}\bra{X_j,P_j}
\label{eq:blockPOVM}
\end{equation} 
which satisfies $\int d\tilde X\,d\tilde P\,\Pi_A(\tilde X,\tilde P)=\Id$.
The associated Kraus operator is taken as $M_A=(2\pi)^{-m/2}\prod_j \ket{X_j,P_j}\bra{X_j,P_j}$.


\begin{figure}[htp]
	\centering
	
	\begin{subfigure}{0.45\textwidth}
		\centering
		\includegraphics[width=\linewidth]{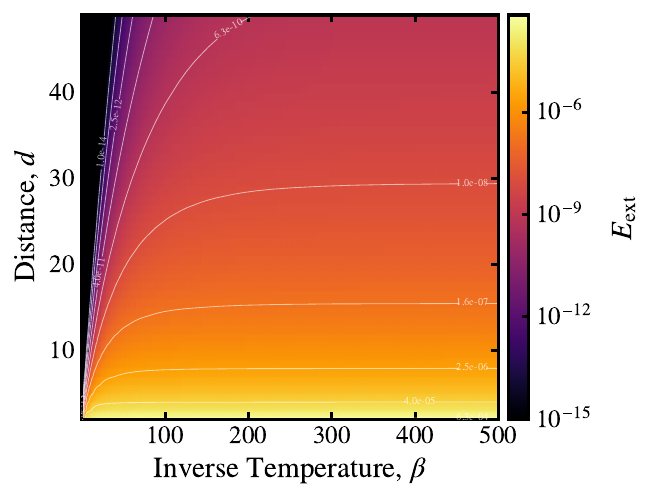}
		\caption{}
		\label{fig:heatmapbd}
	\end{subfigure}
	\hfill
	\begin{subfigure}{0.45\textwidth}
		\centering
		\includegraphics[width=\linewidth]{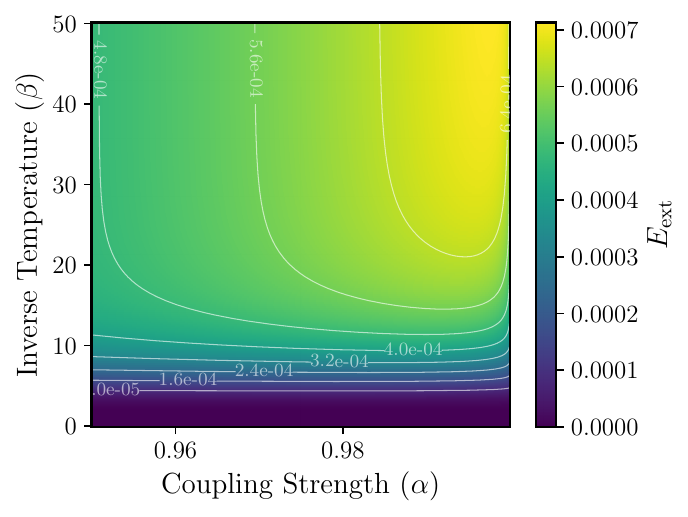}
		\caption{}
		\label{fig:heatmapab}
	\end{subfigure}
	
	\caption{Density plots of the optimized extracted energy
		$E_{\mathrm{ext}}$ in Eq. \eqref{eq:Eextmatrix} as a function of the indicated
		control parameters. Panel (a): dependence on inverse temperature $\beta$ and
		Alice--Bob separation $d$, evaluated at $N=300$, $\alpha=0.9999$, and
		$\Omega=1$. Panel (b): dependence on coupling strength $\alpha$ and inverse
		temperature $\beta$ at fixed separation $d=2$, evaluated at $N=300$ and
		$\Omega=1$.}
	\label{fig:Eext_alpha_beta_distance}
\end{figure}


Alice then transmits the measurement outcomes $(\tilde X,\tilde P)$ to Bob and, upon receiving the measurement outcome, he applies the local displacement
\begin{equation}
U_B(\tilde X,\tilde P)=\exp\left[\ii\left((\bm\theta\cdot \tilde P)q_{n_B}-(\bm\varphi\cdot \tilde X)p_{n_B}\right)\right]
\label{eq:UB}
\end{equation}
where $\bm\theta,\bm\varphi\in\mathbb{R}^m$ are optimized real vectors. To distinguish QET from ordinary energy transport by propagating excitations, Bob must carry out the local operation before the injected energy can propagate from Alice's region to Bob's site.

Because $U_B$ shifts Bob's quadratures as
\begin{equation}
U_B^\dagger p_{n_B}U_B=p_{n_B}+\bm\theta\cdot\tilde P,
\quad
U_B^\dagger q_{n_B}U_B=q_{n_B}+\bm\varphi\cdot\tilde X
\end{equation}
we obtain after averaging over outcomes
\begin{align}
\avg{H_B}_{\mathrm{post}}=&\avg{H_B}_\beta+\frac12\bm\theta^T T_p^{(\beta)}\bm\theta+\big(J_p^{(\beta)}\big)^T\bm\theta
\nonumber\\
&+\frac12\bm\varphi^T T_q^{(\beta)}\bm\varphi+\big(J_q^{(\beta)}\big)^T\bm\varphi
\label{eq:HBpostquadratic}
\end{align}
with
\begin{align}
\big(T_p^{(\beta)}\big)_{jk}&=\avg{p_jp_k}_\beta+\frac{\Omega}{2}\delta_{jk},
&\big(J_p^{(\beta)}\big)_j&=\avg{p_jp_{n_B}}_\beta
\label{eq:TpJp}
\\
\big(T_q^{(\beta)}\big)_{jk}&=\avg{q_jq_k}_\beta+\frac{1}{2\Omega}\delta_{jk},
&\big(J_q^{(\beta)}\big)_j&=\avg{q_jQ_B}_\beta
\label{eq:TqJq}
\end{align}

\amend{The matrices $T_p^{(\beta)}$ and $T_q^{(\beta)}$ are the
	covariance matrices of Alice's classical outcomes $\widetilde
	P$ and $\widetilde X$, respectively. The diagonal terms
	$\Omega/2$ and $1/(2\Omega)$ are the measurement noise added
	by the coherent-state POVM. The vectors $J_p^{(\beta)}$ and
	$J_q^{(\beta)}$ contain the corresponding cross-correlations
	with Bob's momentum $p_{n_B}$ and local force $Q_B$.
	Equation \eqref{eq:Eextmatrix} therefore weights the useful Alice--Bob
	correlations by the inverse covariance of Alice's noisy
	measurement record.}

The baseline term satisfies $\avg{H_B}_{\mathrm{pre}}=\avg{H_B}_{\beta}$, so minimizing Eq. \eqref{eq:HBpostquadratic} gives the optimized extracted energy; Appendix \ref{app:A} contains the full derivation.
\begin{equation}
\Eext(\beta)=\frac12\big(J_p^{(\beta)}\big)^T\big(T_p^{(\beta)}\big)^{-1}J_p^{(\beta)}
+\frac12\big(J_q^{(\beta)}\big)^T\big(T_q^{(\beta)}\big)^{-1}J_q^{(\beta)}
\label{eq:Eextmatrix}
\end{equation}

Equation \eqref{eq:Eextmatrix} is exact for an arbitrary Gaussian block measurement. To obtain analytic control over the temperature and distance dependence, we now specialize to the single-site protocol, where the matrix expression collapses to scalar thermal correlators.

Thus the energy extraction problem reduces to the pre-existing thermal two-point functions. We now use Eq. \eqref{eq:Eextmatrix} to map the extracted energy over the main control parameters. Figure \ref{fig:Eext_alpha_beta_distance}(a) shows the dependence on inverse temperature $\beta$ and separation $d$ in the near-critical regime, while Fig. \ref{fig:Eext_alpha_beta_distance}(b) shows the dependence on coupling $\alpha$ and inverse temperature $\beta$ at fixed separation. These plots reveal the same physical trend from two complementary viewpoints: finite temperature suppresses long-distance extraction, whereas lowering the temperature and approaching criticality both enhance the energy that Bob can extract.

\section{Exact thermodynamic limit\label{sec:exact}}

\subsection{Single-site protocol}

The clearest analytic results arise for a single measured site $A=\{n_A\}$ and a single-site Bob at distance $d=|n_B-n_A|$. In that setting,
\begin{equation}
\Eext(\beta,\alpha;d)=\frac12\frac{h_d(\beta,\alpha)^2}{h_0(\beta,\alpha)+\Omega/2}
+\frac12\frac{J_q(\beta,\alpha;d)^2}{g_0(\beta,\alpha)+1/(2\Omega)}
\label{eq:Eextsingle}
\end{equation}
with
\begin{equation}
J_q(\beta,\alpha;d)=g_d(\beta,\alpha)-\frac\alpha2\Big[g_{d-1}(\beta,\alpha)+g_{d+1}(\beta,\alpha)\Big]
\label{eq:Jqsingle}
\end{equation}
A useful identity holds before any critical limit is taken:
\begin{equation}
J_q(\beta,\alpha;d)=h_d(\beta,\alpha),\quad \alpha<1,\; 0<\beta\le\infty
\label{eq:JqEqhd}
\end{equation}
Indeed, inserting Eq. \eqref{eq:gn_int} into Eq. \eqref{eq:Jqsingle} and using $\cos[(d-1)\theta]+\cos[(d+1)\theta]=2\cos(d\theta)\cos\theta$ yields
\begin{align}
J_q(\beta,\alpha;d)&=\frac{1}{2\pi}\int_0^{2\pi}d\theta\,\frac{1-\alpha\cos\theta}{2\omega_\alpha(\theta)}\coth\!\left(\frac{\beta\omega_\alpha(\theta)}{2}\right)\cos(d\theta)
\nonumber\\
&=\frac{1}{2\pi}\int_0^{2\pi}d\theta\,\frac{\omega_\alpha(\theta)}{2}\coth\!\left(\frac{\beta\omega_\alpha(\theta)}{2}\right)\cos(d\theta)\nonumber\\
&=h_d(\beta,\alpha),
\end{align}
which proves Eq. \eqref{eq:JqEqhd}. This reduction is central because it isolates the entire protocol dependence in a single thermal coupling function. Therefore, equation \eqref{eq:JqEqhd} reduces the single-site problem to the asymptotics of one function, $h_d$. The rest of the analysis therefore asks how this coupling behaves under the ordered thermodynamic and near-critical limits relevant for the harmonic chain.

\section{Large-distance and near-critical asymptotics\label{sec:asymptotics}}

For mathematical consistency, we now enforce the ordered limits appropriate to periodic boundary conditions. One first keeps $\alpha<1$, takes the thermodynamic limit, then chooses the temperature regime, and only afterwards takes the gap-closing limit $\alpha\uparrow 1$. Setting $\alpha=1$ at finite $N$ would make the zero mode singular and would obscure the ordered critical behavior.

\subsection{Fixed finite $\beta$, then $\alpha\uparrow 1$}

For $\beta\in(0,\infty)$ and $\alpha<1$, a Mittag-Leffler expansion of $\coth$ gives
\begin{align}
\frac{1}{2\omega}\coth\!\left(\frac{\beta\omega}{2}\right)
&=\frac{1}{\beta\omega^2}+\frac{2}{\beta}\sum_{m=1}^{\infty}\frac{1}{\omega^2+\nu_m^2},
\label{eq:ML1}
\\
\frac{\omega}{2}\coth\!\left(\frac{\beta\omega}{2}\right)
&=\frac{1}{\beta}+\frac{2\omega^2}{\beta}\sum_{m=1}^{\infty}\frac{1}{\omega^2+\nu_m^2},
\label{eq:ML2}
\end{align}
with bosonic Matsubara frequencies $\nu_m=2\pi m/\beta$. The remaining cosine integrals are rational and can be evaluated by contour integration. Defining
$a_m:=1+\nu_m^2,\quad\Delta_m(\alpha):=\sqrt{a_m^2-\alpha^2},\quad
r_m(\alpha):=\frac{a_m-\Delta_m(\alpha)}{\alpha}$, and for the $m=0$ term of $g_n$, $\Delta_0(\alpha):=\sqrt{1-\alpha^2},
\quad r_0(\alpha):=\frac{1-\Delta_0(\alpha)}{\alpha},$
one finds
\begin{align}
g_n(\beta,\alpha)&=\frac{1}{\beta}\frac{r_0(\alpha)^n}{\Delta_0(\alpha)}
+\frac{2}{\beta}\sum_{m=1}^{\infty}\frac{r_m(\alpha)^n}{\Delta_m(\alpha)},
\label{eq:gnfinitebetaexact}
\\
h_n(\beta,\alpha)&=-\frac{2}{\beta}\sum_{m=1}^{\infty}\frac{\nu_m^2}{\Delta_m(\alpha)}r_m(\alpha)^n,
\quad n\ge 1,
\label{eq:hnfinitebetaexact}
\\
h_0(\beta,\alpha)&=\frac{1}{\beta}+\frac{2}{\beta}\sum_{m=1}^{\infty}\left(1-\frac{\nu_m^2}{\Delta_m(\alpha)}\right).
\label{eq:h0finitebetaexact}
\end{align}

The derivation is given in Appendix \ref{app:B}.

In the ordered critical limit $\alpha\uparrow 1$, the onsite position correlator diverges as $g_0(\beta,\alpha)\sim \frac{1}{\beta\sqrt{1-\alpha^2}},\quad \alpha\uparrow 1,$
whereas $h_d$ remains finite. Hence the $q$-sector contribution in Eq. \eqref{eq:Eextsingle} vanishes. Writing $\kappa_m:=\operatorname{arsinh}\!\left(\frac{\nu_m}{\sqrt2}\right)=\operatorname{arsinh}\!\left(\frac{\sqrt2\pi m}{\beta}\right),
$
so that $\tanh\kappa_m=\nu_m/\sqrt{\nu_m^2+2}$ and $r_m^{\crit}=\ee^{-2\kappa_m}$, we obtain
\begin{align}
h_d^{\crit}(\beta)&=-\frac{2}{\beta}\sum_{m=1}^{\infty}\tanh\kappa_m\,\ee^{-2d\kappa_m},
\quad d\ge 1
\label{eq:hdcritfinitebeta}
\\
h_0^{\crit}(\beta)&=\frac{1}{\beta}+\frac{2}{\beta}\sum_{m=1}^{\infty}\big(1-\tanh\kappa_m\big)
\label{eq:h0critfinitebeta}
\end{align}

Therefore
\begin{equation}
\Eext^{\crit}(\beta;d)=\frac12\frac{\big(h_d^{\crit}(\beta)\big)^2}{h_0^{\crit}(\beta)+\Omega/2},
\quad 0<\beta<\infty
\label{eq:Ecritfinitebeta}
\end{equation}
At large $d$ the $m=1$ term dominates, so $h_d^{\crit}(\beta)\sim -\frac{2}{\beta}\tanh\kappa_1\,\ee^{-2d\kappa_1},$
and thus
\begin{equation}
\Eext^{\crit}(\beta;d)\sim
\frac{2\tanh^2\kappa_1}{\beta^2\left[h_0^{\crit}(\beta)+\Omega/2\right]}\,\ee^{-4d\kappa_1}
\label{eq:ElargeDbeta}
\end{equation}
The thermal correlation length controlling $h_d$ is therefore
\begin{equation}
\xi_\beta=\frac{1}{2\kappa_1}=
\left[2\operatorname{arsinh}\!\left(\frac{\sqrt2\pi}{\beta}\right)\right]^{-1}
\label{eq:xibeta}
\end{equation}

\amend{Here $\xi_\beta$ is the thermal correlation length of
	$h_d^{\mathrm{crit}}(\beta)$. Since Eq. \eqref{eq:Ecritfinitebeta} is quadratic in
	this correlator, the extracted energy decays on the shorter
	length $\xi_E=\xi_\beta/2=(4\kappa_1)^{-1}$.}

\begin{figure}[!htp]
	\includegraphics[width=0.9\columnwidth]{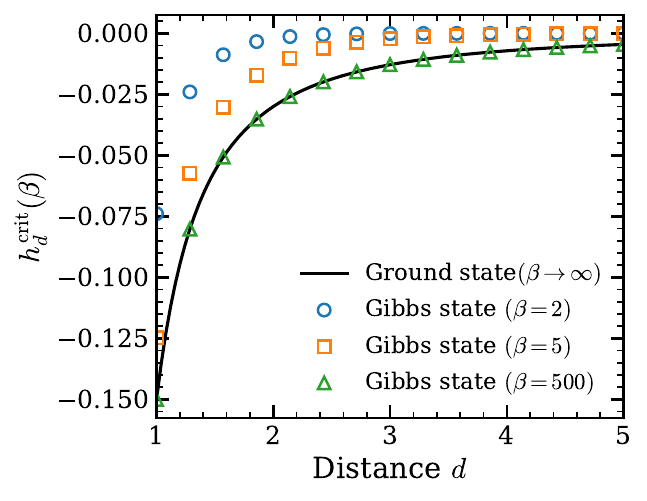}
	\caption{Critical thermal coupling $h_d^{\mathrm{crit}}(\beta)$ in the thermodynamic limit. Symbols: finite-temperature results from Eq. \eqref{eq:hdcritfinitebeta} at selected values of $\beta$. Solid line: exact zero-temperature result from Eq. \eqref{eq:hdcritzeroint}. At finite $\beta$, thermal screening suppresses the coupling at large distance; as $\beta\to\infty$ the curves converge to the ground-state profile.}
	\label{fig:hd_crititcal}
\end{figure}

Before turning to the zero-temperature limit, we examine the critical thermal coupling $h_d^{\mathrm{crit}}(\beta)$ that enters the single-site protocol directly. Figure \ref{fig:hd_crititcal} plots this quantity in the thermodynamic critical limit for several values of $\beta$, alongside the exact ground-state result. Two features are immediately apparent. First, $|h_d^{\mathrm{crit}}(\beta)|$ decays rapidly with distance at any finite 
temperature, consistent with the thermal screening length set by Eq. \eqref{eq:xibeta}. Second, as $\beta$ increases the finite temperature curves converge toward the ground state profile, anticipating the algebraic long distance behavior derived in the following subsection.

\subsection{$\beta\to\infty$, then $\alpha\uparrow 1$}

At zero temperature one first sets $\coth(\beta\omega/2)\to 1$ at fixed $\alpha<1$. The position correlator again diverges in the ordered critical limit, now logarithmically, and the $q$ sector vanishes (see Appendix B). For the momentum coupling one can evaluate the critical integral exactly:
\begin{equation}
h_d^{\crit}(\infty)=\frac{1}{2\pi}\int_0^{2\pi}d\theta\,\frac{\sqrt{1-\cos\theta}}{2}\cos(d\theta).
\label{eq:hdcritzeroint}
\end{equation}
Using $\sqrt{1-\cos\theta}=\sqrt2\sin(\theta/2)$ on $[0,2\pi]$ gives $h_0^{\crit}(\infty)=\frac{\sqrt2}{\pi},\quad
h_d^{\crit}(\infty)=-\frac{\sqrt2}{\pi(4d^2-1)},\quad d\ge 1$,
Hence
\begin{equation}
\Eext^{\crit}(\infty;d)=\frac{1}{\pi^2(4d^2-1)^2\left(\sqrt2/\pi+\Omega/2\right)}
\label{eq:Ecritzero}
\end{equation}
For large separation, $h_d^{\crit}(\infty)\sim -\frac{\sqrt2}{4\pi}d^{-2}$. Therefore,
\begin{equation}
\Eext^{\crit}(\infty;d)\sim
\frac{1}{16\pi^2\left(\sqrt2/\pi+\Omega/2\right)}d^{-4}.
\label{eq:dminus4}
\end{equation}
This resolves the long-distance critical law analytically.

\begin{figure}[!htp]
	\includegraphics[width=0.9\columnwidth]{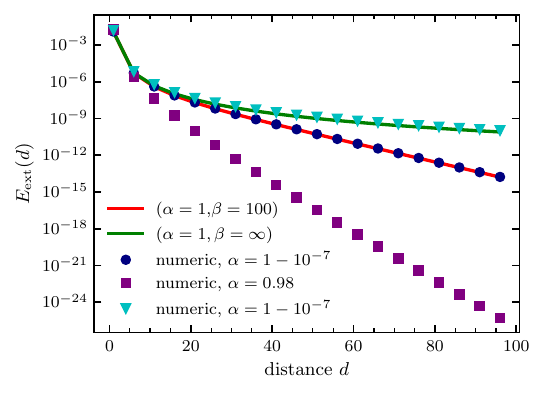}
	\caption{Extracted energy $E_{\mathrm{ext}}(d)$ in the thermodynamic regime. Solid curves: analytical critical results ($\alpha=1$) for the Gibbs state at finite temperature and for the ground state. Symbols: numerical data for $N=2^{15}$ in the near-critical regime ($\alpha=1-10^{-7}$, circles) and in the gapped phase ($\alpha=0.98$, squares). The near-critical data are nearly indistinguishable from the $\alpha=1$ curves at this scale, whereas the gapped data exhibit a substantially faster decay.}
	\label{fig:Eext_benchmark}
\end{figure}

We benchmark the asymptotic formulas against thermodynamic-limit numerics across the critical, near-critical, and gapped regimes. 
Figure \ref{fig:Eext_benchmark} compares analytical critical curves at $\alpha=1$ with numerical data from a finite chain of $N=2^{15}$ sites. The near-critical data at $\alpha=1-10^{-7}$ track the critical predictions closely over the full distance range, confirming that the critical formulas remain accurate even slightly away from the exact critical point. In contrast, the gapped ground-state data at $\alpha=0.98$ decay far more steeply, as expected once a finite correlation length enters the picture. Notably, finite-temperature QET at criticality can exceed gapped ground-state QET over a broad window of separations, indicating that proximity to a critical point 
can outweigh the thermal penalty. The residual deviation between the near-critical numerical points and the exact analytical curves is indiscernible at this scale and is examined separately in Fig. \ref{fig:d4E_comparison} of Appendix \ref{app:B}.

\section{Crossover Diagnostics\label{sec:numerics}}

The asymptotic formulas above identify the limiting decay
laws. We next show how these regimes appear in finite-size
numerics and how the apparent exponent crosses over before
the true asymptotic behavior becomes visible.

\amend{The physical distinction between these regimes is set by
	the gap and thermal correlation lengths. From the factor
	$r_m^d$ in Eq. \eqref{eq:hnfinitebetaexact}, one has
	$r_m=e^{-\operatorname{arcosh}[(1+\nu_m^2)/\alpha]}$.
	At fixed finite $\beta$, the $m=1$ term has the longest
	range and defines
	$\xi_h(\alpha,\beta)=
	\{\operatorname{arcosh}[(1+\nu_1^2)/\alpha]\}^{-1}$, where
	$\nu_1=2\pi/\beta$. In the ordered critical limit this
	reduces to the thermal length $\xi_\beta$ in Eq. \eqref{eq:xibeta},
	whereas the zero-temperature limit at fixed $\alpha<1$
	gives the gap length
	$\xi_\Delta=[\operatorname{arcosh}(1/\alpha)]^{-1}$.
	Using the low-momentum dispersion in Eq. \eqref{eq:hn_int}, the latter
	satisfies $\xi_\Delta\simeq v/\Delta_{\mathrm{gap}}$ near
	criticality, with
	$\Delta_{\mathrm{gap}}=\sqrt{1-\alpha}$ and
	$v=\sqrt{\alpha/2}$. Thus a nonzero gap limits the range of
	ground-state correlations, while finite temperature
	introduces an additional cutoff through the lowest
	Matsubara frequency. The near-critical
	low-temperature regime has both $\xi_\Delta$ and
	$\xi_\beta$ larger than the displayed distances; the
	near-critical finite-temperature regime is instead limited
	mainly by $\xi_\beta$, while the gapped ground state is
	limited by $\xi_\Delta$. For the gapped ground-state regime, this behavior is qualitatively consistent with general exponential decaying bound on QET energy extraction in finite-range gapped lattice systems \cite{HAQUE2026131613}.}
	
\begin{figure}[htp]
\centering
\includegraphics[width=0.91\textwidth]{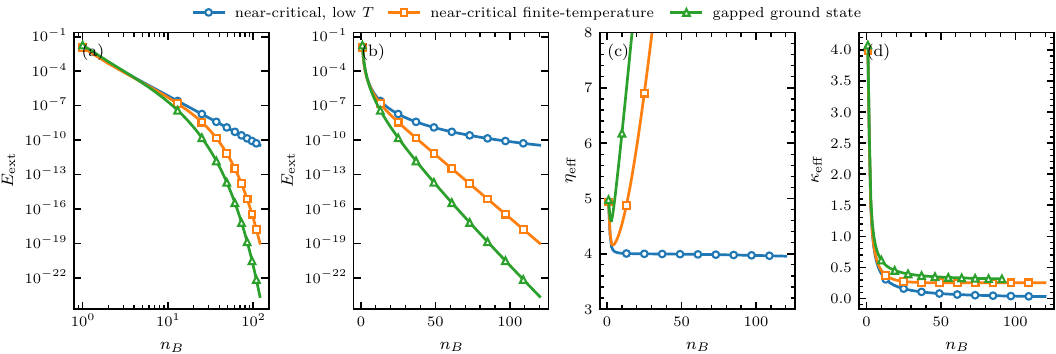}
	\caption{
		\amend{Distance-scaling diagnostics for the extracted energy
		$E_{\mathrm{ext}}(n_B)$. The three curves correspond to:
		near-critical low temperature (blue),
		$(N,\alpha,\beta,\Omega)
		=(2000,1-10^{-9},
		10^5,1)$;
		near-critical finite temperature (orange),
		$(N,\alpha,\beta,\Omega)
		=(2000,1-10^{-9}, 70, 1)$;
		and the gapped ground-state regime (green),
		$(N,\alpha,\beta,\Omega)
		=(2000, 0.99, 10^5, 1)$.
		Panel (a) shows the data on logarithmic axes, testing
		algebraic decay, while panel (b) shows the same data on
		semilogarithmic axes, testing exponential decay.
		Panel (c) shows the effective logarithmic exponent
		$\eta_{\mathrm{eff}}(n_B)
		=-d\log E_{\mathrm{ext}}/d\log n_B$, and panel (d) shows
		the local attenuation rate
		$\kappa_{\mathrm{eff}}(n_B)
		=-d\log E_{\mathrm{ext}}/dn_B$.}}
		
	\label{fig:Eext_distance_scaling}
\end{figure}

	Figure \ref{fig:Eext_distance_scaling} examines the spatial decay of the extracted energy $E_{\rm ext}(n_B\footnote{to avoid confusion between derivative and distance, we used $n_B:=d$})$ for three representative regimes: a near critical low temperature case, a near critical finite temperature case, and a gapped ground state case. The distinction between algebraic and exponential suppression is most transparent from (a) and (b). In the near critical low-temperature regime, the curve is closest to linear on the log-log scale over an extended intermediate range of $n_B$, consistent with an approximate algebraic decay before eventual finite correlation length effects set in. By contrast, the finite temperature near critical case and the gapped ground state bend strongly on the log-log plot and are substantially closer to linear on the semi-log plot, which is the expected signature of exponential suppression at large separation. (c) and (d) provide local diagnostics of this behavior. The effective logarithmic slope $\eta_{\rm eff}(n_B)$ remains nearly constant for the near critical low-temperature case over the accessible window, whereas it grows steadily in the thermally suppressed and gapped cases. Similarly, the local attenuation rate $\kappa_{\rm eff}(n_B)$ tends toward a small value in the near critical low-temperature regime, but approaches a finite nonzero scale in the other two cases, consistent with exponential decay. \amend{More explicitly, the numerics agree with the critical form
		$E_{\mathrm{ext}}\propto n_B^{-4}$ from Eq. \eqref{eq:dminus4} and gives
		$\eta_{\mathrm{eff}}\simeq4$ and
		$\kappa_{\mathrm{eff}}\simeq4/n_B\to0$. An exponential
		envelope $E_{\mathrm{ext}}\propto
		\exp(-n_B/\xi_E)$ instead gives
		$\eta_{\mathrm{eff}}\simeq n_B/\xi_E$ and
		$\kappa_{\mathrm{eff}}\simeq1/\xi_E$. Hence the growing
		$\eta_{\mathrm{eff}}$ and finite
		$\kappa_{\mathrm{eff}}$ in the thermally screened and
		gapped cases directly measure the finite length that cuts
		off the critical correlations.}		
\onecolumngrid

\twocolumngrid

\section{Squeezed Gaussian POVM\label{sec:squeezing}}
Since the large-distance behavior is fixed by the thermal correlators, the natural next question is whether Gaussian measurement design can improve the extraction prefactor without changing the scaling law. We therefore examine squeezed coherent-state POVMs.

The coherent-state POVM is not the only Gaussian choice. Let Alice instead measure squeezed coherent states with squeeze parameter $r\ge0$ and squeeze angle $\vartheta$. The only change relevant to the averaged energy is the measurement noise added to the second moments. For a $q$-squeezed measurement, $\vartheta=0$ (see Appendix \ref{app:C}),
\begin{equation}
\nu_X(r)=\frac{\ee^{-2r}}{2\Omega},
\quad
\nu_P(r)=\frac{\Omega\ee^{2r}}{2},
\label{eq:qxnoise}
\end{equation}
whereas for a $p$-squeezed measurement, $\vartheta=\pi/2$,
\begin{equation}
\nu_X(r)=\frac{\ee^{2r}}{2\Omega},
\quad
\nu_P(r)=\frac{\Omega\ee^{-2r}}{2}.
\label{eq:pxnoise}
\end{equation}
The optimized single-site extracted energy becomes
\begin{equation}
\Eext=\frac12\frac{h_d(\beta,\alpha)^2}{h_0(\beta,\alpha)+\nu_P(r,\vartheta)}
+\frac12\frac{J_q(\beta,\alpha;d)^2}{g_0(\beta,\alpha)+\nu_X(r,\vartheta)}.
\label{eq:Eextsq}
\end{equation}

\begin{center}
	\includegraphics[width=\columnwidth]{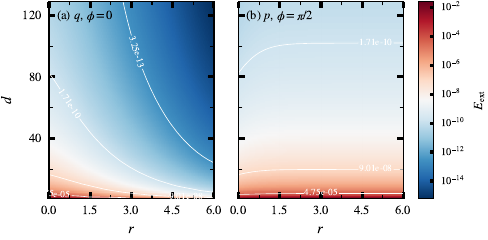}
	\captionof{figure}{Density plots of the extracted energy $E_{\mathrm{ext}}$ versus squeezing parameter $r$ and separation $d$ for $N_{\mathrm{sys}}=500$, $\beta=10^{5}$, $\Omega=1.0$, and $\alpha=1-10^{-13}$. Panel (a) corresponds to the $q$-squeezed measurement, and panel (b) to the $p$-squeezed measurement. The white lines indicate contours of constant extracted energy.}
	\label{fig:fig5}
\end{center}

In the ordered critical limit the $q$ sector again drops out, so the only surviving effect of squeezing is the replacement $\Omega/2\mapsto \nu_P$ in the denominator. Thus $p$ squeezing lowers the effective momentum noise and enhances the extracted energy, while $q$ squeezing raises it and suppresses the yield. Figure \ref{fig:fig5} illustrates this point exactly. Importantly, the numerators are untouched. Squeezing therefore changes the efficiency but not the large-distance scaling. This is also visible in Fig. \ref{fig:sPOVM_loglog}. 

\onecolumngrid
\begin{center}
		\includegraphics[width=\textwidth]{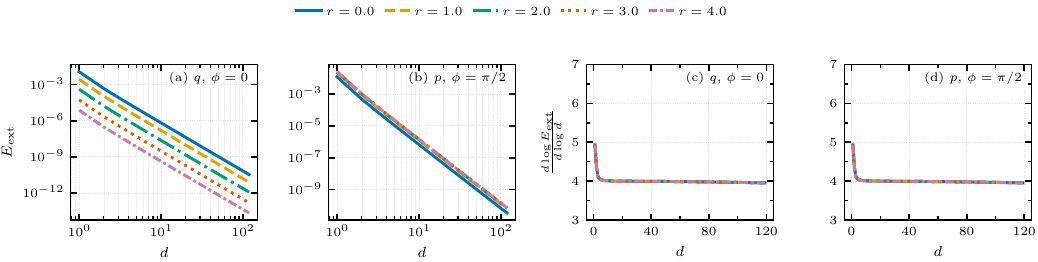}
		\captionof{figure}{Distance dependence of the extracted energy $E_{\mathrm{ext}}$ for squeezed Gaussian measurements at zero temperature. The system parameters are $N_{\mathrm{sys}}=2000$, $\beta=10^{5}$, $\Omega=1.0$, and $\alpha=1-10^{-13}$. Panels (a) and (b) show $E_{\mathrm{ext}}$ versus the separation $d$ on a log-log scale for different squeezing parameters $r=0,1,2,3,4$. Panel (a) corresponds to the $q$-squeezed measurement, while panel (b) corresponds to the $p$-squeezed measurement. Panels (c) and (d) display the local logarithmic slope, $d\log E_{\mathrm{ext}}/d\log d$, for the same two measurement choices. In both cases the large-distance behavior approaches the same asymptotic exponent, close to $-4$, while squeezing mainly changes the overall extraction amplitude.}
		\label{fig:sPOVM_loglog}
\end{center}

\twocolumngrid

\section{Discussion and conclusion\label{sec:discussion}}

In this work we have placed quantum energy teleportation in the harmonic chain on a controlled finite-temperature footing. For translationally invariant Gibbs states, the optimized Gaussian protocol reduces to thermal two-point functions, and in the single-site setting the full distance dependence collapses to one correlator, $h_d$. This reduction makes the ordered limits transparent. At fixed finite temperature, the leading Matsubara contribution sets a thermal correlation length $\xi_\beta$ and produces exponential suppression of the extracted energy. In the zero-temperature critical limit taken after the thermodynamic limit, the coupling admits a closed form, which yields the asymptotic law $\Eext\sim d^{-4}$.

These results sharpen the status of the harmonic chain as an analytically tractable QET platform. In the present setting, the protocol is controlled less by protocol-specific details than by the correlation structure of the underlying Gaussian state. Finite temperature screens the useful correlations, while proximity to criticality restores long-range tails. The squeezed-measurement analysis supports the same conclusion: changing the Gaussian measurement modifies the effective noise and therefore the extraction efficiency, but it does not alter the large-distance scaling class.

The present analysis is limited to Gaussian measurements and primarily to single-site extraction. Natural extensions include finite-temperature block protocols and multisite feedback on Bob's side, which may improve the extraction prefactor while preserving the same asymptotic scaling. Because the same quadratic Hamiltonian also describes coupled microwave resonators after canonical rescaling, the present results may also serve as a benchmark for bosonic QET implementations in circuit-QED architectures.

\begin{acknowledgments}
I would like to thank Professor Stefan Kehrein for suggesting the original idea behind this work and for his guidance and helpful discussions during my MSc project. I am also grateful to him for his encouragement and support to write this paper.
\end{acknowledgments}

\section*{Data Availability}
The code supporting the findings of this study is openly available in \doi{10.5281/zenodo.19648101}.

\onecolumngrid
\appendix

\section{Derivation of the Gaussian block-measurement formula}
\label{app:A}

In this appendix we derive the optimized extracted-energy formula quoted in Eq. \eqref{eq:Eextmatrix} of the main text for an arbitrary Gaussian block measurement. The derivation proceeds in three steps. We first establish the coherent-state POVM identities needed for outcome averaging. We then compute the transformed local energy under Bob's conditional displacement. Finally, we minimize the resulting quadratic form over Bob's feedback parameters.

\subsection{Coherent-state POVM on Alice's block}

Alice measures a block $A$ consisting of $m=2\ell+1$ sites. On each measured site she introduces the operator
\begin{equation}
	\hat b=\sqrt{\frac{\Omega}{2}}\,\hat q+\frac{i}{\sqrt{2\Omega}}\,\hat p,
	\qquad
	\Omega>0.
	\label{appA:bdef}
\end{equation}
The corresponding coherent states $|X,P\rangle$ are defined by the eigenvalue equation
\begin{equation}
	\hat b\,|X,P\rangle=c\,|X,P\rangle,
	\qquad
	c=\sqrt{\frac{\Omega}{2}}X+\frac{i}{\sqrt{2\Omega}}P.
	\label{appA:coh_eig}
\end{equation}
The one-site POVM element is
\begin{equation}
	\Pi(X,P)=\frac{1}{2\pi}|X,P\rangle\langle X,P|.
	\label{appA:single_povm}
\end{equation}

To verify completeness, we begin from the standard coherent-state identity
\begin{equation}
	\int\frac{d^2c}{\pi}\,|c\rangle\langle c|=\mathbb{I}.
	\label{appA:coh_resolution}
\end{equation}
Writing $c=u+iv$ and comparing with Eq. \eqref{appA:coh_eig}, we obtain
\begin{equation}
	u=\sqrt{\frac{\Omega}{2}}X,
	\qquad
	v=\frac{1}{\sqrt{2\Omega}}P.
	\label{appA:uv_coords}
\end{equation}
Hence the Jacobian is
\begin{equation}
	du\,dv=
	\sqrt{\frac{\Omega}{2}}\frac{1}{\sqrt{2\Omega}}\,dX\,dP
	=
	\frac{1}{2}\,dX\,dP.
	\label{appA:jacobian}
\end{equation}
Substituting Eq. \eqref{appA:jacobian} into Eq. \eqref{appA:coh_resolution}, we find
\begin{equation}
	\int dX\,dP\,\Pi(X,P)=\mathbb{I}.
	\label{appA:single_complete}
\end{equation}

For the full block measurement we define
\begin{equation}
	\Pi_A(\tilde X,\tilde P)=\frac{1}{(2\pi)^m}\prod_{j=1}^{m}|X_j,P_j\rangle\langle X_j,P_j|,
	\label{appA:block_povm}
\end{equation}
where $\tilde X=(X_1,\dots,X_m)$ and $\tilde P=(P_1,\dots,P_m)$. Since the one-site POVM is complete, the block POVM satisfies
\begin{equation}
	\int d\tilde X\,d\tilde P\,\Pi_A(\tilde X,\tilde P)=\mathbb{I}.
	\label{appA:block_complete}
\end{equation}
The associated Kraus operator is taken as
\begin{equation}
	M_A(\tilde X,\tilde P)=\frac{1}{(\sqrt{2\pi})^m}\prod_{j=1}^{m}|X_j,P_j\rangle\langle X_j,P_j|,
	\label{appA:block_kraus}
\end{equation}
so that
\begin{equation}
	M_A^\dagger M_A=\Pi_A.
	\label{appA:kraus_povm}
\end{equation}

\subsection{First and second moments of the POVM}

To evaluate the averaged local energy, we require the first and second moments of the measurement outcomes. These moments can be derived from the coherent-state identities
\begin{equation}
	\int\frac{d^2c}{\pi}\,c\,|c\rangle\langle c|=\hat b,
	\qquad
	\int\frac{d^2c}{\pi}\,c^\ast\,|c\rangle\langle c|=\hat b^\dagger.
	\label{appA:c_moments}
\end{equation}
The proof is standard. Expanding $|c\rangle$ in the number basis and using Gaussian integration gives the matrix elements of $\hat b$ and $\hat b^\dagger$ directly.

Using Eq. \eqref{appA:coh_eig}, we may rewrite the classical variables as
\begin{equation}
	X=\frac{c+c^\ast}{\sqrt{2\Omega}},
	\qquad
	P=-i\sqrt{\frac{\Omega}{2}}(c-c^\ast).
	\label{appA:XP_from_c}
\end{equation}
Substituting Eq. \eqref{appA:c_moments} into Eq. \eqref{appA:XP_from_c}, we obtain the one-site first moments
\begin{equation}
	\int dX\,dP\,X\,\Pi(X,P)=q,
	\qquad
	\int dX\,dP\,P\,\Pi(X,P)=p.
	\label{appA:first_moments}
\end{equation}

The second moments follow similarly. Using
\begin{equation}
	X^2=\frac{c^2+(c^\ast)^2+2|c|^2}{2\Omega},
	\qquad
	P^2=-\frac{\Omega}{2}(c-c^\ast)^2,
	\label{appA:quadratic_XP}
\end{equation}
together with the standard identity
\begin{equation}
	\int\frac{d^2c}{\pi}\,|c|^2\,|c\rangle\langle c|=\hat b\hat b^\dagger,
	\label{appA:abs_c_squared}
\end{equation}
one obtains
\begin{equation}
	\int dX\,dP\,X^2\,\Pi(X,P)=q^2+\frac{1}{2\Omega}\mathbb{I},
	\qquad
	\int dX\,dP\,P^2\,\Pi(X,P)=p^2+\frac{\Omega}{2}\mathbb{I}.
	\label{appA:second_moments}
\end{equation}

Because the block POVM factorizes across Alice's sites, the multisite moments are immediate:
\begin{equation}
	\int d\tilde X\,d\tilde P\,X_j\,\Pi_A=q_j,
	\qquad
	\int d\tilde X\,d\tilde P\,P_j\,\Pi_A=p_j,
	\label{appA:block_first_moments}
\end{equation}
and
\begin{equation}
	\int d\tilde X\,d\tilde P\,X_jX_k\,\Pi_A
	=
	q_jq_k+\frac{1}{2\Omega}\delta_{jk}\mathbb{I},
	\qquad
	\int d\tilde X\,d\tilde P\,P_jP_k\,\Pi_A
	=
	p_jp_k+\frac{\Omega}{2}\delta_{jk}\mathbb{I}.
	\label{appA:block_second_moments}
\end{equation}

\subsection{Bob's conditional displacement}

After receiving the classical outcome $(\tilde X,\tilde P)$, Bob applies the local displacement
\begin{equation}
	U_B(\tilde X,\tilde P)=
	\exp\!\left[
	i\Bigl((\theta\cdot\tilde P)q_{n_B}-(\phi\cdot\tilde X)p_{n_B}\Bigr)
	\right],
	\label{appA:UB}
\end{equation}
where $\theta,\phi\in\mathbb{R}^m$ are real control vectors. For fixed outcome, define
\begin{equation}
	a:=\theta\cdot\tilde P=\sum_{j=1}^{m}\theta_jP_j,
	\qquad
	b:=\phi\cdot\tilde X=\sum_{j=1}^{m}\phi_jX_j.
	\label{appA:a_b}
\end{equation}
Then Eq. \eqref{appA:UB} becomes $U_B=\exp[i(aq_{n_B}-bp_{n_B})]$.

Applying the Baker-Campbell-Hausdorff formula and noting that the relevant nested commutators vanish after the first step, we obtain
\begin{equation}
	U_B^\dagger p_{n_B}U_B=p_{n_B}+a,
	\qquad
	U_B^\dagger q_{n_B}U_B=q_{n_B}+b.
	\label{appA:quadrature_shifts}
\end{equation}
It is convenient to introduce the operator
\begin{equation}
	Q_B:=q_{n_B}-\frac{\alpha}{2}(q_{n_B-1}+q_{n_B+1}),
	\label{appA:QB}
\end{equation}
since this is precisely the linear combination appearing in the local energy around Bob. Substituting Eq. \eqref{appA:quadrature_shifts} into Eq. (12) of the main text, we find
\begin{equation}
	U_B^\dagger H_B U_B
	=
	H_B+a\,p_{n_B}+\frac{1}{2}a^2+b\,Q_B+\frac{1}{2}b^2.
	\label{appA:HB_shifted}
\end{equation}

\subsection{Outcome averaging and quadratic form}

Alice's unnormalized post-measurement state is
\begin{equation}
	\rho(\tilde X,\tilde P)=M_A(\tilde X,\tilde P)\rho_\beta M_A(\tilde X,\tilde P).
	\label{appA:post_meas_state}
\end{equation}
The averaged pre-Bob and post-Bob states are
\begin{equation}
	\rho_{\mathrm{pre}}
	=
	\int d\tilde X\,d\tilde P\,M_A\rho_\beta M_A,
	\qquad
	\rho_{\mathrm{post}}(\theta,\phi)
	=
	\int d\tilde X\,d\tilde P\,U_BM_A\rho_\beta M_AU_B^\dagger.
	\label{appA:pre_post_states}
\end{equation}

Now let $O$ be any operator supported outside Alice's measured block. Since $[O,\Pi_A]=0$, cyclicity of the trace gives
\begin{equation}
	\Tr(M_A\rho_\beta M_A\,O)=\Tr(\rho_\beta\,O\,\Pi_A).
	\label{appA:trace_identity}
\end{equation}
Applying Eq. \eqref{appA:trace_identity} to $O=U_B^\dagger H_BU_B$ and using Eqs. \eqref{appA:block_first_moments}-\eqref{appA:block_second_moments}, we obtain
\begin{align}
	\langle H_B\rangle_{\mathrm{post}}(\theta,\phi)
	&=
	\langle H_B\rangle_\beta
	+\sum_{j=1}^{m}\theta_j\langle p_jp_{n_B}\rangle_\beta
	+\frac{1}{2}\sum_{j,k=1}^{m}\theta_j
	\left(
	\langle p_jp_k\rangle_\beta+\frac{\Omega}{2}\delta_{jk}
	\right)\theta_k
	\nonumber\\
	&\quad
	+\sum_{j=1}^{m}\phi_j\langle q_jQ_B\rangle_\beta
	+\frac{1}{2}\sum_{j,k=1}^{m}\phi_j
	\left(
	\langle q_jq_k\rangle_\beta+\frac{1}{2\Omega}\delta_{jk}
	\right)\phi_k.
	\label{appA:HB_post_quad}
\end{align}

Introducing the matrices and vectors already used in the main text,
\begin{equation}
	\bigl(T_p^{(\beta)}\bigr)_{jk}:=\langle p_jp_k\rangle_\beta+\frac{\Omega}{2}\delta_{jk},
	\qquad
	\bigl(T_q^{(\beta)}\bigr)_{jk}:=\langle q_jq_k\rangle_\beta+\frac{1}{2\Omega}\delta_{jk},
	\label{appA:TpTq}
\end{equation}
\begin{equation}
	\bigl(J_p^{(\beta)}\bigr)_j:=\langle p_jp_{n_B}\rangle_\beta,
	\qquad
	\bigl(J_q^{(\beta)}\bigr)_j:=\langle q_jQ_B\rangle_\beta,
	\label{appA:JpJq}
\end{equation}
Eq. \eqref{appA:HB_post_quad} becomes
\begin{equation}
	\langle H_B\rangle_{\mathrm{post}}(\theta,\phi)
	=
	\langle H_B\rangle_\beta
	+\frac{1}{2}\theta^TT_p^{(\beta)}\theta
	+\bigl(J_p^{(\beta)}\bigr)^T\theta
	+\frac{1}{2}\phi^TT_q^{(\beta)}\phi
	+\bigl(J_q^{(\beta)}\bigr)^T\phi.
	\label{appA:HB_post_matrix}
\end{equation}

\subsection{Optimization over Bob's feedback parameters}

Since $T_p^{(\beta)}$ and $T_q^{(\beta)}$ are positive definite, Eq. \eqref{appA:HB_post_matrix} is minimized at the stationary point. Differentiating with respect to $\theta$ and $\phi$ gives
$$
T_p^{(\beta)}\theta_{\mathrm{opt}}+J_p^{(\beta)}=0,
\qquad
T_q^{(\beta)}\phi_{\mathrm{opt}}+J_q^{(\beta)}=0.
$$
Hence
\begin{equation}
	\theta_{\mathrm{opt}}=-\bigl(T_p^{(\beta)}\bigr)^{-1}J_p^{(\beta)},
	\qquad
	\phi_{\mathrm{opt}}=-\bigl(T_q^{(\beta)}\bigr)^{-1}J_q^{(\beta)}.
	\label{appA:theta_phi_opt}
\end{equation}
Substituting these expressions back into Eq. \eqref{appA:HB_post_matrix}, we obtain
\begin{equation}
	\langle H_B\rangle_{\mathrm{post,opt}}
	=
	\langle H_B\rangle_\beta
	-\frac{1}{2}\bigl(J_p^{(\beta)}\bigr)^T\bigl(T_p^{(\beta)}\bigr)^{-1}J_p^{(\beta)}
	-\frac{1}{2}\bigl(J_q^{(\beta)}\bigr)^T\bigl(T_q^{(\beta)}\bigr)^{-1}J_q^{(\beta)}.
	\label{appA:HB_post_opt}
\end{equation}

If Bob does nothing, then $\theta=\phi=0$, and therefore $\langle H_B\rangle_{\mathrm{pre}}=\langle H_B\rangle_\beta$. Comparing with the definition of extracted energy in the main text, the baseline term cancels, and we recover Eq. (21):
\begin{equation}
	E_{\mathrm{ext}}(\beta)
	=
	\frac{1}{2}\bigl(J_p^{(\beta)}\bigr)^T\bigl(T_p^{(\beta)}\bigr)^{-1}J_p^{(\beta)}
	+
	\frac{1}{2}\bigl(J_q^{(\beta)}\bigr)^T\bigl(T_q^{(\beta)}\bigr)^{-1}J_q^{(\beta)}.
	\label{appA:Eext_general}
\end{equation}

\section{Thermodynamic correlators and ordered critical limits}
\label{app:B}

In this Appendix we derive the thermodynamic correlators used in Eqs. \eqref{eq:gnfinitebetaexact}-\eqref{eq:h0finitebetaexact} of the main text and then analyze their ordered critical limits. The crucial point is that for periodic boundary conditions the gap-closing limit $\alpha\uparrow1$ must be taken only after the thermodynamic limit.

\subsection{Normal modes and Gibbs covariances}

Introduce the Fourier angles
\begin{equation}
	\theta_k=\frac{2\pi k}{N},
	\qquad
	k=0,1,\dots,N-1,
	\label{appB:theta_k}
\end{equation}
and the Fourier modes
\begin{equation}
	q_j=\frac{1}{\sqrt{N}}\sum_{k=0}^{N-1}Q_k e^{i\theta_k j},
	\qquad
	p_j=\frac{1}{\sqrt{N}}\sum_{k=0}^{N-1}P_k e^{i\theta_k j}.
	\label{appB:fourier_modes}
\end{equation}
Applying the periodic boundary condition to Eq. \eqref{appB:fourier_modes}, we find that the Hamiltonian diagonalizes as
\begin{equation}
	H=\frac{1}{2}\sum_{k=0}^{N-1}\Bigl(P_kP_{-k}+\omega_k^2Q_kQ_{-k}\Bigr),
	\qquad
	\omega_k^2=1-\alpha\cos\theta_k.
	\label{appB:H_diagonal}
\end{equation}
For $\alpha<1$, all $\omega_k$ are strictly positive, including the zero mode. At $\alpha=1$, however, the zero mode becomes gapless and the periodic Gibbs state is no longer trace class. This is why the order of limits matters.

Define the ladder operators
$$
a_k=\sqrt{\frac{\omega_k}{2}}Q_k+\frac{i}{\sqrt{2\omega_k}}P_k,
\qquad
a_k^\dagger=\sqrt{\frac{\omega_k}{2}}Q_{-k}-\frac{i}{\sqrt{2\omega_k}}P_{-k}.
$$
Then
$$
H=\sum_{k=0}^{N-1}\omega_k\left(a_k^\dagger a_k+\frac{1}{2}\right).
$$
In the Gibbs state the Bose occupation number is
$$
n_k=\langle a_k^\dagger a_k\rangle_{\beta,\alpha}=\frac{1}{e^{\beta\omega_k}-1},
\qquad
n_k+\frac{1}{2}=\frac{1}{2}\coth\!\left(\frac{\beta\omega_k}{2}\right).
$$
Therefore the nonvanishing mode covariances are
$$
\langle Q_kQ_{-k}\rangle_{\beta,\alpha}
=
\frac{1}{2\omega_k}\coth\!\left(\frac{\beta\omega_k}{2}\right),
\qquad
\langle P_kP_{-k}\rangle_{\beta,\alpha}
=
\frac{\omega_k}{2}\coth\!\left(\frac{\beta\omega_k}{2}\right).
$$

\subsection{Thermodynamic correlators}

Define the translation-invariant correlators
\begin{equation}
	g_n(\beta,\alpha;N):=\langle q_jq_{j+n}\rangle_{\beta,\alpha},
	\qquad
	h_n(\beta,\alpha;N):=\langle p_jp_{j+n}\rangle_{\beta,\alpha}.
	\label{appB:gn_hn_finiteN}
\end{equation}
Substituting the Fourier expansion and using the thermal covariances above yields
\begin{equation}
	g_n(\beta,\alpha;N)
	=
	\frac{1}{N}\sum_{k=0}^{N-1}
	\frac{1}{2\omega_k}\coth\!\left(\frac{\beta\omega_k}{2}\right)\cos(n\theta_k),
	\label{appB:gn_finiteN}
\end{equation}
and
\begin{equation}
	h_n(\beta,\alpha;N)
	=
	\frac{1}{N}\sum_{k=0}^{N-1}
	\frac{\omega_k}{2}\coth\!\left(\frac{\beta\omega_k}{2}\right)\cos(n\theta_k).
	\label{appB:hn_finiteN}
\end{equation}

Now fix $\alpha<1$ and take the thermodynamic limit. The Riemann sums become
\begin{equation}
	g_n(\beta,\alpha)
	=
	\frac{1}{2\pi}\int_0^{2\pi}d\theta\,
	\frac{1}{2\omega_\alpha(\theta)}
	\coth\!\left(\frac{\beta\omega_\alpha(\theta)}{2}\right)\cos(n\theta),
	\label{appB:gn_thermo}
\end{equation}
and
\begin{equation}
	h_n(\beta,\alpha)
	=
	\frac{1}{2\pi}\int_0^{2\pi}d\theta\,
	\frac{\omega_\alpha(\theta)}{2}
	\coth\!\left(\frac{\beta\omega_\alpha(\theta)}{2}\right)\cos(n\theta),
	\label{appB:hn_thermo}
\end{equation}
where
\begin{equation}
	\omega_\alpha(\theta)=\sqrt{1-\alpha\cos\theta}.
	\label{appB:omega_alpha}
\end{equation}

\subsection{Single-site reduction and the identity $J_q=h_d$}

For the single-site protocol discussed in Sec. IV of the main text, the extracted energy takes the scalar form
\begin{equation}
	E_{\mathrm{ext}}(\beta,\alpha;d)
	=
	\frac{1}{2}\frac{h_d(\beta,\alpha)^2}{h_0(\beta,\alpha)+\Omega/2}
	+
	\frac{1}{2}\frac{J_q(\beta,\alpha;d)^2}{g_0(\beta,\alpha)+1/(2\Omega)},
	\label{appB:single_site_energy}
\end{equation}
with
\begin{equation}
	J_q(\beta,\alpha;d)
	=
	g_d(\beta,\alpha)-\frac{\alpha}{2}\Bigl(g_{d-1}(\beta,\alpha)+g_{d+1}(\beta,\alpha)\Bigr).
	\label{appB:Jq_def}
\end{equation}
A central identity is
\begin{equation}
	J_q(\beta,\alpha;d)=h_d(\beta,\alpha),
	\qquad
	\alpha<1,\quad 0<\beta\le\infty.
	\label{appB:Jq_eq_hd}
\end{equation}
To prove it, define
$$
A_{\beta,\alpha}(\theta):=
\frac{1}{2\omega_\alpha(\theta)}
\coth\!\left(\frac{\beta\omega_\alpha(\theta)}{2}\right).
$$
Then Eq. \eqref{appB:gn_thermo} reads
$$
g_n(\beta,\alpha)=\frac{1}{2\pi}\int_0^{2\pi}d\theta\,A_{\beta,\alpha}(\theta)\cos(n\theta).
$$
Substituting this expression into Eq. \eqref{appB:Jq_def} and using
$$
\cos((d-1)\theta)+\cos((d+1)\theta)=2\cos(d\theta)\cos\theta,
$$
we obtain
$$
J_q
=
\frac{1}{2\pi}\int_0^{2\pi}d\theta\,
A_{\beta,\alpha}(\theta)\cos(d\theta)\bigl(1-\alpha\cos\theta\bigr).
$$
Since $1-\alpha\cos\theta=\omega_\alpha(\theta)^2$, the integrand collapses immediately, yielding Eq. \eqref{appB:Jq_eq_hd}.

\subsection{Ordered limits}

For periodic boundary conditions we enforce the limit order
\begin{equation}
	\alpha<1,
	\qquad
	N\to\infty,
	\qquad
	\text{choose the temperature regime},
	\qquad
	\alpha\uparrow1\ \text{last}.
	\label{appB:ordered_limits}
\end{equation}
This ordering is mandatory. If one sets $\alpha=1$ before taking $N\to\infty$, the zero mode becomes singular and the Gibbs state ceases to be well defined.

\subsection{Fixed finite temperature, then $\alpha\uparrow1$}

Fix $0<\beta<\infty$. To evaluate the thermodynamic integrals exactly, we use the Mittag-Leffler expansion
\begin{equation}
	\coth x=\frac{1}{x}+2x\sum_{m=1}^{\infty}\frac{1}{x^2+\pi^2m^2}.
	\label{appB:mittag_leffler}
\end{equation}
Setting $x=\beta\omega/2$ and introducing the bosonic Matsubara frequencies
\begin{equation}
	\nu_m=\frac{2\pi m}{\beta},
	\qquad
	m\ge1,
	\label{appB:matsubara}
\end{equation}
we obtain
\begin{equation}
	\frac{1}{2\omega}\coth\!\left(\frac{\beta\omega}{2}\right)
	=
	\frac{1}{\beta\omega^2}
	+\frac{2}{\beta}\sum_{m=1}^{\infty}\frac{1}{\omega^2+\nu_m^2},
	\label{appB:kernel_qq}
\end{equation}
and
\begin{equation}
	\frac{\omega}{2}\coth\!\left(\frac{\beta\omega}{2}\right)
	=
	\frac{1}{\beta}
	+\frac{2\omega^2}{\beta}\sum_{m=1}^{\infty}\frac{1}{\omega^2+\nu_m^2}.
	\label{appB:kernel_pp}
\end{equation}

The remaining angular integrals reduce to the rational cosine integral
$$
J_n(a,\alpha):=
\frac{1}{2\pi}\int_0^{2\pi}\frac{\cos(n\theta)}{a-\alpha\cos\theta}\,d\theta,
\qquad
a>\alpha>0.
$$
Passing to the unit circle via $z=e^{i\theta}$ and evaluating the residue of the pole inside $|z|=1$, one finds
\begin{equation}
	J_n(a,\alpha)
	=
	\frac{1}{\sqrt{a^2-\alpha^2}}
	\left(
	\frac{a-\sqrt{a^2-\alpha^2}}{\alpha}
	\right)^n.
	\label{appB:J_n_closed}
\end{equation}

Define
$$
a_m:=1+\nu_m^2,
\qquad
\Delta_m(\alpha):=\sqrt{a_m^2-\alpha^2},
\qquad
r_m(\alpha):=\frac{a_m-\Delta_m(\alpha)}{\alpha},
$$
and for the $m=0$ contribution in $g_n$,
$$
\Delta_0(\alpha):=\sqrt{1-\alpha^2},
\qquad
r_0(\alpha):=\frac{1-\Delta_0(\alpha)}{\alpha}.
$$
Substituting Eq. \eqref{appB:J_n_closed} into the thermodynamic correlators yields
\begin{equation}
	g_n(\beta,\alpha)
	=
	\frac{1}{\beta}\frac{r_0(\alpha)^n}{\Delta_0(\alpha)}
	+
	\frac{2}{\beta}\sum_{m=1}^{\infty}\frac{r_m(\alpha)^n}{\Delta_m(\alpha)},
	\label{appB:gn_exact}
\end{equation}
\begin{equation}
	h_n(\beta,\alpha)
	=
	-\frac{2}{\beta}\sum_{m=1}^{\infty}
	\frac{\nu_m^2}{\Delta_m(\alpha)}\,r_m(\alpha)^n,
	\qquad
	n\ge1,
	\label{appB:hn_exact}
\end{equation}
and
\begin{equation}
	h_0(\beta,\alpha)
	=
	\frac{1}{\beta}
	+\frac{2}{\beta}\sum_{m=1}^{\infty}
	\left(
	1-\frac{\nu_m^2}{\Delta_m(\alpha)}
	\right).
	\label{appB:h0_exact}
\end{equation}
These are exactly the formulas we have used in Eqs. (28)-(30) of the main text.

We now take the ordered critical limit. The divergence of $g_0$ is dominated by the $m=0$ term:
$$
g_0(\beta,\alpha)\supset\frac{1}{\beta\sqrt{1-\alpha^2}}
\longrightarrow+\infty
\qquad
(\alpha\uparrow1).
$$
Therefore the $q$-sector of Eq. \eqref{appB:single_site_energy} vanishes, while the numerator remains finite because of Eq. \eqref{appB:Jq_eq_hd}.

To write the surviving correlator compactly, define
\begin{equation}
	\kappa_m:=\operatorname{arsinh}\!\left(\frac{\nu_m}{\sqrt2}\right)
	=
	\operatorname{arsinh}\!\left(\frac{\sqrt2\pi m}{\beta}\right).
	\label{appB:kappa_m}
\end{equation}
Then
$$
\tanh\kappa_m=\frac{\nu_m}{\sqrt{\nu_m^2+2}},
\qquad
e^{-2\kappa_m}=
\frac{1}{1+\nu_m^2+\nu_m\sqrt{\nu_m^2+2}}.
$$
Hence the critical momentum correlators are
\begin{equation}
	h_d^{\mathrm{crit}}(\beta)
	=
	-\frac{2}{\beta}\sum_{m=1}^{\infty}\tanh(\kappa_m)e^{-2d\kappa_m},
	\qquad
	d\ge1,
	\label{appB:hd_crit_beta}
\end{equation}
and
\begin{equation}
	h_0^{\mathrm{crit}}(\beta)
	=
	\frac{1}{\beta}
	+\frac{2}{\beta}\sum_{m=1}^{\infty}\bigl(1-\tanh\kappa_m\bigr).
	\label{appB:h0_crit_beta}
\end{equation}
Substituting these into Eq. \eqref{appB:single_site_energy}, we obtain
\begin{equation}
	E_{\mathrm{ext}}^{\mathrm{crit}}(\beta;d)
	=
	\frac{1}{2}
	\frac{\bigl(h_d^{\mathrm{crit}}(\beta)\bigr)^2}
	{h_0^{\mathrm{crit}}(\beta)+\Omega/2},
	\qquad
	0<\beta<\infty.
	\label{appB:Eext_crit_beta}
\end{equation}

At large separation the $m=1$ term dominates the sum. Therefore
$$
h_d^{\mathrm{crit}}(\beta)\sim
-\frac{2}{\beta}\tanh(\kappa_1)e^{-2d\kappa_1},
$$
and the extracted energy decays as
\begin{equation}
	E_{\mathrm{ext}}^{\mathrm{crit}}(\beta;d)
	\sim
	\frac{2\tanh^2(\kappa_1)}
	{\beta^2\bigl[h_0^{\mathrm{crit}}(\beta)+\Omega/2\bigr]}
	e^{-4d\kappa_1}.
	\label{appB:finite_beta_asymptotic}
\end{equation}
Thus the useful energy is exponentially screened at any fixed finite temperature.

\subsection{Zero temperature, then $\alpha\uparrow1$}

We now set $\beta=\infty$ at fixed $\alpha<1$. In this case the hyperbolic cotangent reduces to unity, so
$$
g_n(\infty,\alpha)
=
\frac{1}{2\pi}\int_0^{2\pi}\frac{\cos(n\theta)}{2\omega_\alpha(\theta)}\,d\theta,
\qquad
h_n(\infty,\alpha)
=
\frac{1}{2\pi}\int_0^{2\pi}\frac{\omega_\alpha(\theta)}{2}\cos(n\theta)\,d\theta.
$$

The position variance again diverges in the ordered critical limit, now logarithmically. Writing $\mu^2:=1-\alpha$, one has for small $\theta$
$$
\omega_\alpha(\theta)\sim\sqrt{\mu^2+\theta^2/2},
$$
and therefore
$$
g_0(\infty,\alpha)\sim\int_0^\Lambda\frac{d\theta}{\sqrt{\mu^2+\theta^2/2}}
\sim \ln\frac{1}{\mu}.
$$
Thus the $q$ sector in Eq. \eqref{appB:single_site_energy} again drops out.

For the momentum correlator, however, dominated convergence allows one to set $\alpha=1$ directly in the integrand:
\begin{equation}
	h_d^{\mathrm{crit}}(\infty)
	=
	\frac{1}{2\pi}\int_0^{2\pi}
	\frac{\sqrt{1-\cos\theta}}{2}\cos(d\theta)\,d\theta.
	\label{appB:hd_crit_infty_integral}
\end{equation}
Using
$$
\sqrt{1-\cos\theta}=\sqrt2\sin\!\left(\frac{\theta}{2}\right),
\qquad
0\le\theta\le2\pi,
$$
we obtain
$$
h_d^{\mathrm{crit}}(\infty)
=
\frac{\sqrt2}{4\pi}
\int_0^{2\pi}\sin\!\left(\frac{\theta}{2}\right)\cos(d\theta)\,d\theta.
$$
Expanding the product with $\sin a\cos b=\frac{1}{2}[\sin(a+b)+\sin(a-b)]$ and integrating term by term yields
$$
\int_0^{2\pi}\sin\!\left(\frac{\theta}{2}\right)\cos(d\theta)\,d\theta
=
-\frac{4}{4d^2-1}.
$$
Therefore
\begin{equation}
	h_0^{\mathrm{crit}}(\infty)=\frac{\sqrt2}{\pi},
	\qquad
	h_d^{\mathrm{crit}}(\infty)=
	-\frac{\sqrt2}{\pi(4d^2-1)},
	\qquad
	d\ge1.
	\label{appB:hd_crit_infty_closed}
\end{equation}
Substituting this into Eq. \eqref{appB:single_site_energy}, we recover Eq. (37) of the main text:
\begin{equation}
	E_{\mathrm{ext}}^{\mathrm{crit}}(\infty;d)
	=
	\frac{1}{\pi^2(4d^2-1)^2\bigl(\sqrt2/\pi+\Omega/2\bigr)}.
	\label{appB:Eext_crit_infty}
\end{equation}
At large separation this becomes
\begin{equation}
	E_{\mathrm{ext}}^{\mathrm{crit}}(\infty;d)
	\sim
	\frac{1}{16\pi^2\bigl(\sqrt2/\pi+\Omega/2\bigr)}\,d^{-4},
	\label{appB:power_law}
\end{equation}
which is the analytic $d^{-4}$ law stated in the main text.

\begin{figure}[h]
	\centering
	\includegraphics[width=0.5\textwidth]{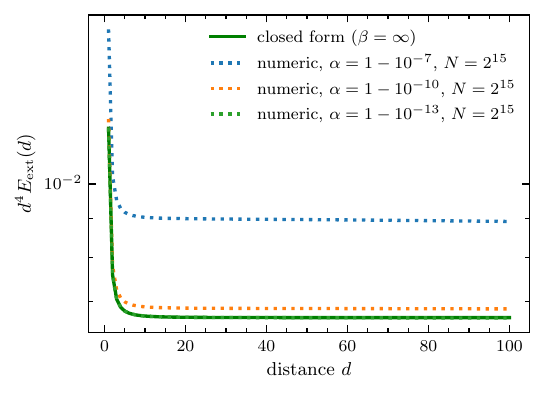}
	\caption{Comparison of the rescaled quantity $d^4E_{\mathrm{ext}}(d)$ with the closed-form zero-temperature critical prediction. The finite-size numerical curves approach the analytic result as $\alpha$ is tuned closer to unity, consistent with the asymptotic behavior $E_{\mathrm{ext}}(d)\sim Cd^{-4}$.}
	\label{fig:d4E_comparison}
\end{figure}

As a numerical check of the zero-temperature asymptotics, Fig. \ref{fig:d4E_comparison} plots the rescaled quantity $d^4E_{\mathrm{ext}}(d)$. If the large-distance law is indeed $E_{\mathrm{ext}}(d)\sim Cd^{-4}$, this rescaled combination should approach a constant at sufficiently large $d$. The figure shows precisely this behavior. The analytic $\beta=\infty$ closed form is compared with finite-$N$ evaluations performed at progressively smaller gaps, $\alpha=1-10^{-7}$, $1-10^{-10}$, and $1-10^{-13}$, all for $N=2^{15}$. As $\alpha$ is tuned closer to unity, the numerical curves move toward the closed-form prediction over an extended range of distances, confirming that the ordered limit is captured correctly. In this way the plot serves as a direct benchmark of the analytic coefficient in the $d^{-4}$ tail, rather than merely of the exponent.

\section{Squeezed Gaussian POVM}
\label{app:C}

For completeness we also record the squeezed-POVM extension underlying Sec. VII of the main text. The only effect of squeezing is to modify the measurement-added noise in the second moments.
Let $S(\zeta)$ be the single-mode squeeze operator with $\zeta=re^{i\vartheta}$ and $r\ge0$. The squeezed coherent states are
\begin{equation}
	|X,P;\zeta\rangle:=D(X,P)S(\zeta)|0\rangle,
	\qquad
	D(X,P)=e^{iPq-iXp},
	\label{appD:squeezed_states}
\end{equation}
and the one-site POVM element is
\begin{equation}
	\Pi_\zeta(X,P)=\frac{1}{2\pi}|X,P;\zeta\rangle\langle X,P;\zeta|.
	\label{appD:squeezed_POVM}
\end{equation}
For the block measurement one takes the product over Alice's measured sites.

For a $q$-squeezed measurement, $\vartheta=0$, the second moments become
\begin{equation}
	\nu_X(r)=\frac{e^{-2r}}{2\Omega},
	\qquad
	\nu_P(r)=\frac{\Omega e^{2r}}{2},
	\label{appD:q_squeezed_noise}
\end{equation}
whereas for a $p$-squeezed measurement, $\vartheta=\pi/2$,
\begin{equation}
	\nu_X(r)=\frac{e^{2r}}{2\Omega},
	\qquad
	\nu_P(r)=\frac{\Omega e^{-2r}}{2}.
	\label{appD:p_squeezed_noise}
\end{equation}
Therefore the covariance matrices in Appendix \ref{app:A} are replaced by
\begin{equation}
	\bigl(T_p^{(\beta,\zeta)}\bigr)_{jk}
	=
	\langle p_jp_k\rangle_\beta+\nu_P(r)\delta_{jk},
	\qquad
	\bigl(T_q^{(\beta,\zeta)}\bigr)_{jk}
	=
	\langle q_jq_k\rangle_\beta+\nu_X(r)\delta_{jk},
	\label{appD:squeezed_covariances}
\end{equation}
while the correlation vectors remain unchanged. Repeating the same Gaussian minimization as in Appendix A, one finds
\begin{equation}
	E_{\mathrm{ext}}(\beta;r,\vartheta)
	=
	\frac{1}{2}\bigl(J_p^{(\beta)}\bigr)^T\bigl(T_p^{(\beta,\zeta)}\bigr)^{-1}J_p^{(\beta)}
	+
	\frac{1}{2}\bigl(J_q^{(\beta)}\bigr)^T\bigl(T_q^{(\beta,\zeta)}\bigr)^{-1}J_q^{(\beta)}.
	\label{appD:Eext_squeezed_general}
\end{equation}
For the single-site protocol this reduces to
\begin{equation}
	E_{\mathrm{ext}}(\beta,\alpha;d;r,\vartheta)
	=
	\frac{1}{2}\frac{h_d(\beta,\alpha)^2}{h_0(\beta,\alpha)+\nu_P(r,\vartheta)}
	+
	\frac{1}{2}
	\frac{
		\Bigl[g_d(\beta,\alpha)-\frac{\alpha}{2}\bigl(g_{d-1}(\beta,\alpha)+g_{d+1}(\beta,\alpha)\bigr)\Bigr]^2
	}
	{g_0(\beta,\alpha)+\nu_X(r,\vartheta)}.
	\label{appD:Eext_squeezed_single}
\end{equation}
In the ordered critical limit the $q$ sector again vanishes, so the only surviving effect of squeezing is the replacement
$$
\frac{\Omega}{2}\longmapsto \nu_P(r,\vartheta)
$$
in the denominator of the momentum-sector contribution. Consequently,
\begin{equation}
	E_{\mathrm{ext}}^{\mathrm{crit}}(\beta;d;r,\vartheta)
	=
	\frac{1}{2}
	\frac{\bigl(h_d^{\mathrm{crit}}(\beta)\bigr)^2}
	{h_0^{\mathrm{crit}}(\beta)+\nu_P(r,\vartheta)}.
	\label{appD:Eext_squeezed_crit}
\end{equation}
Thus squeezing changes the extraction prefactor, but not the large-distance scaling class. 
\twocolumngrid
\bibliography{qet_prb_refs}

\end{document}